\documentstyle{paper}
\begin{document}
\newcommand{\Df}[2]{\mbox{$\frac{#1}{#2}$}}

\hfill  INP-96-10/417, hep-ph/9603267

\centerline{Explicit solutions of the 3--loop vacuum integral 
recurrence relations}

\centerline{\footnotesize
P.\,A.\,BAIKOV\footnote{
Supported in part
by the Russian Basic Research Foundation (grant N 93--02--14428),
INTAS (grant 91-1180) and by Japan Society to Promote Science. \\ 
\hspace*{3pt}Email: baikov@theory.npi.msu.su}}
\vspace*{0.015truein}
\centerline{\footnotesize\it Institute of Nuclear Physics,
Moscow State University}
\baselineskip=10pt   
\centerline{\footnotesize\it  119~899, Moscow, Russia}

\vspace{3mm}
\centerline{
\begin{minipage}{14.5cm}
Explicit formulas for the solutions of the recurrence 
relations for 3--loop vacuum integrals are suggested.
This formulas can be used for direct calculations and demonstrate a
high efficiency.
They also produce a new type of recurrence
relations over the space--time dimension.
\end{minipage}}

\section{Introduction}
Recurrence relations  are powerful tools for
evaluating multi--loop Feynman integrals \cite{ch-tk}.
They relate Feynman integrals with various
degrees of their denominators. In many cases
they provide the possibility to express an
integral with given degrees of denominators as a linear
combination of a few master integrals with some coefficients which
we will call weight factors.

At two--loop level the recurrence relations are relatively simple and one
can easily find and realize the corresponding recursive algorithm for
the calculation of the weight factors.

At three--loop level the recurrence relations are more complicated and to 
find an algorithm to calculate weight factors is a serious problem.

For vacuum integrals with one nonzero mass and various numbers of 
massless lines the corresponding
algorithm was described in 
\cite{REC}-\cite{av-p}.
Here a repeated application of various recurrence relations is performed
until the integrals of a certain type are eliminated. 

In recent works \cite{REC},\cite{AFMT}-\cite{chet} the recursive algorithm
was successfully applied in calculations of
two--loop and three--loop QED and QCD.

Nevertheless, such recursive algorithms
lead to too time and memory consuming calculations
because of the size of intermediate
expressions grows exponentially with respect to the degrees of
the denominators in the initial integral.  
In fact, the calculations mentioned above were made at the
limits of computer capabilities.

In this work we suggest a new approach based on explicit formulas for 
the solutions of the recurrence relations. As an application, the case
of three loop vacuum integrals with four equal mass and two
massless lines is considered. The efficiency of this approach
is demonstrated by calculations of previousely unknown coefficients in 
Taylor expansion of QED photon vacuum polarization for small $q^2$.

\section{General case}

Let us consider the three--loop vacuum integrals with six different masses:

\begin{eqnarray}
B(\underline{n},D)\equiv
B(n_1,n_2,n_3,n_4,n_5,n_6,D)=
\frac{m^{2\Sigma_1^6 n_i-3D}}
{\big[\imath\pi^{D/2}\Gamma(3-D/2)\big]^3}
\int\int\int \frac{d^Dp\,d^Dk\,d^Dl} 
{D_1^{n_1}D_2^{n_2}D_3^{n_3}D_4^{n_4}D_5^{n_5}D_6^{n_6}}
\label{integral}
\end{eqnarray}

\noindent
where

\centerline{
\begin{tabular}{lll}
$D_1=k^2-\mu_1 m^2$,&
$D_2=l^2-\mu_2 m^2$,&
$D_3=(p+k)^2-\mu_3 m^2$\\
$D_4=(p+l)^2-\mu_4 m^2$,&
$D_5=(p+k+l)^2-\mu_5 m^2$,&
$D_6=p^2-\mu_6 m^2$\\
\end{tabular}
}

Let us derive recurrence relations that result from integration by parts,
by letting  $(\partial/\partial p_i)\cdot p_j$ act on
the integrand, with $p_{i,j}\in\{p,k,l\}$. For example, 
acting by $(\partial/\partial k)\cdot (p+k)$ we get

\begin{eqnarray}
(D-2n_3-n_1-n_5) \}B(\underline{n},D)&=&
\{ n_1 {\bf 1}^+({\bf 3}^- -{\bf 6}^- + \mu_3 -\mu_6 +\mu_1)
+2n_3 {\bf 3}^+ \mu_3 \nonumber\\
&&+n_5 {\bf 5}^+({\bf 3}^- -{\bf 2}^- + \mu_3 -\mu_2 +\mu_5)\}B(\underline{n},D)
\label{rr}
\end{eqnarray}

\noindent
where  
${\bf 1}^\pm B(n_1,\ldots)\equiv B(n_1\pm1,\ldots)$, etc.

Others relations can be obtained from (\ref{rr}) by proper 
permutations of the $n_i, \mu_i$ and ${\bf I}^\pm$ objects. 

The common way of using these relations is step by step re-expression
of the integral (\ref{integral}) with some values of $n_i$ through a set of 
integrals 
with shifted values of $n_i$, with the final goal to reduce this set to
a few integrals with $n_i$ are equal to $0$ or $1$, so called "master" 
integrals. The result can be represented as

\begin{eqnarray}
B(\underline{n},D)=\sum_k f^k(\underline{n},D)N_k(D)\nonumber
\end{eqnarray}

\noindent
where the index $k$ enumerate master integrals $N_k(D)$ and corresponding 
coefficient functions $f^k(n_i,D)$. 

There are two problems on this way. First, there is no general 
approach to construction of such recursive procedure, that is to find 
proper combinations of these relations and a proper sequence of its use 
is the matter of art even for the case of one mass 
\cite{av-p}. Second, even in cases when such procedures were 
constructed, they lead to very time and memory consuming calculation
because of large reproduction rate at every recursion step.
For example, the relation (\ref{rr}) expresses the integral through 7 
others.

Instead, let us construct the coefficient functions $f^k(\underline{n},D)$ 
directly as solutions of the given recurrence relations. 

For that, let us diagonalize the recurrence relations with respect to 
$n_i\,{\bf I}^+$ operators. 
We found that the recurrence relations can be represented in the following 
simple form

\begin{eqnarray}
\{P(x_1,\dots,x_6)\cdot n_i{\bf I}^+ - 
\frac{D-4}{2}\partial_i(P(x_1,\dots,x_6)) \}_{x_i={\bf\small I}^- 
+\mu_i} B(\underline{n},D)=0,\quad i=1,\dots, 6.
\label{rr2}
\end{eqnarray}

\noindent
where  
\begin{eqnarray}
P(x_1,\dots,x_6)&=&
2(x_1x_2(x_1+x_2)+x_3x_4(x_3+x_4)+x_5x_6(x_5+x_6))\nonumber\\
&&+x_1x_3x_6+x_1x_4x_5+x_2x_3x_5+x_2x_4x_6\nonumber\\
&&-(x_1+x_2+x_3+x_4+x_5+x_6)(x_1x_2+x_3x_4+x_5x_6)\nonumber
\end{eqnarray}

The differential equation corresponding to (\ref{rr2}) has the solution
$P^{D/2-2}(x_i+\mu_i)$. Let
us consider "Lourent" coefficients of this function:

\begin{eqnarray}
f(n_i,D)=
\frac{1}{(2\pi\imath)^6}
\oint\oint\oint\oint\oint\oint
\frac
{dx_1dx_2dx_3dx_4dx_5dx_6}
{x_1^{n_1}x_2^{n_2}x_3^{n_3}x_4^{n_4}x_5^{n_5}x_6^{n_6}}
{P(x_1+\mu_1,\dots,x_6+\mu_6)^{D/2-2}}
\label{solution}
\end{eqnarray}

\noindent
where integral symbols denote six subsequent complex 
integrations with contours 
which will be described below. If one acts by (\ref{rr2}) on 
(\ref{solution}) one gets up to the surface terms 
the same expression
as acting by $P\partial_i -(D/2-2)(\partial_iP)$ on $P^{D/2-2}$, that is 
zero. Then, the surface terms can be removed if we choose
closed or ended in infinity point contours. For more accuracy one can 
consider 
analytical continuations of the result on $D$ from large negative values.
So (\ref{solution}) is the solution of the relations (\ref{rr}),
and the different choices of the contours correspond to different 
independent solutions. Note, that if one chooses the contour as a small 
circle over zero, one get the true Lourent coefficient
of the function $P^{D/2-2}$, so this function can be called generalized 
generating function for the solutions of the relations (\ref{rr}).

Then, in accordance with the dimensional regularization rules, 
the integrals (\ref{integral}) are not equal to zero only if at least three
among $n_i$ are positive. So it is natural to construct 
the solutions from those that are equal to zero if the index from 
definite three--index set ("Taylor" indexes) is not positive. 
One can obtain such solutions 
if one chooses the contours, corresponding to these indexes,
as circles over zero. In this case these three integrations can be
evaluated and lead to coefficient in the common Taylor expansion in
corresponding variables.

The three remaining integrations in general case lead to the sum of 
generalized hypergeometric seria, but for some cases of practical interest
(see below)
can be reduced to the finite sums of Pochhammers symbols products. 

\section{Example}

As an example let us consider the case of integrals with four equal mass
and two massless lines, that is $\mu_1=\mu_2=0,\mu_3=\mu_4=\mu_5=\mu_6=1$. 
Let us calculate the coefficient functions 
which corresponds to the choice of the master integrals from \cite{REC}.
That is, we expand $B(\underline{n})$ as

\begin{eqnarray}
B(\underline{n},D)=
N(\underline{n},D)B(0,0,1,1,1,1,D)+
M(\underline{n},D)B(1,1,0,0,1,1,D)+
T(\underline{n},D)B(0,0,0,1,1,1,D)\nonumber
\end{eqnarray}

\noindent
with the following normalization conditions
\begin{eqnarray}
N(0,0,1,1,1,1,D)=1,\quad N(1,1,0,0,1,1,D)=0,\quad 
N(0,0,0,1,1,1,D)=0,\label{condN}\\
M(0,0,1,1,1,1,D)=0,\quad M(1,1,0,0,1,1,D)=1,\quad
M(0,0,0,1,1,1,D)=0,\label{condM}\\
T(0,0,1,1,1,1,D)=0,\quad T(1,1,0,0,1,1,D)=0,\quad 
T(0,0,0,1,1,1,D)=1,\label{condT} \end{eqnarray}

The practical rule for choosing the integration contours is: circle around
zero for unity in the master integral and contour over cut for zero in 
the master integral.

To get $N(\underline{n})$ one should make first the Taylor expansion 
in $x_3,x_4,x_5,x_6$

\begin{eqnarray}
B(n_i,D)\propto\oint\oint
\frac{dx_1dx_2}{x_1^{n_1}x_2^{n_2}}
\big(\frac{\partial_3^{n_3-1}\dots\partial_6^{n_6-1}}
{(n_3-1)!\dots(n_6-1)!}
P(x_1,x_2,x_3+1,\dots,x_6+1)^{D/2-2}\big)
\vert_{x_3,\dots,x_6=0}\nonumber
\end{eqnarray}

The remaining integrals over $x_1,x_2$ are of the type

\begin{eqnarray}
\oint\oint
\frac{dx_1dx_2}{x_1^{n_1}x_2^{n_2}}
[x_1x_2(x_1+x_2-4)]^{D/2-2}
\propto
(-4)^{-n_1-n_2}\frac{(D/2-1)_{-n_1}(D/2-1)_{-n_2}}{(3D/2-3)_{-n_1-n_2}}
\equiv N(n_1,n_2,1,1,1,1,D)\nonumber
\end{eqnarray}

\noindent
where we follow the normalization (\ref{condN}). 

The case $M(\underline{n},D)$ is analogous. The only difference is that 
due to the 
symmetry of the task we should take the sum of the solutions with the 
signatures $(++\pm\pm++)$ and $(++++\pm\pm)$.

The case $T(\underline{n},D)$ is more complicated. The symmetry of the task 
assumes that one should try the following form of the solution

\begin{eqnarray}
T(n_1,n_2,n_3,n_4,n_5,n_6,D)&=&
\phantom{+}t(n_1,n_2,n_3,n_4,n_5,n_6,D)+t(n_1,n_2,n_4,n_3,n_6,n_5,D)\nonumber\\
&&+t(n_1,n_2,n_5,n_6,n_3,n_4,D)+t(n_1,n_2,n_6,n_5,n_4,n_3,D)
\nonumber
\end{eqnarray}

\noindent
where $t(\underline{n},D)$ is non--zero only if $n_4,n_5,n_6>0$.
Let us construct $t(\underline{n},D)$ using (\ref{solution}),
keeping in mind possible mixing with $N(\underline{n},D)$ solution.
After differentiating 
over last three indexes the task reduces to the construction of 
$t(n_1,n_2,n_3,1,1,1,D)$. Let us consider the corresponding integral:

\begin{eqnarray}
\overline{t}(n_1,n_2,n_3,D)=
\frac{1}{(2\pi\imath)^3}
\oint\oint\oint
\frac
{dx_1dx_2dx_3}
{x_1^{n_1}x_2^{n_2}x_3^{n_3}}
{(x_3^2-x_1x_2x_3+x_1x_2(x_1+x_2-4))^{D/2-2}}
\label{tbar}
\end{eqnarray}

For $n_3<1$ one can calculates this integral immediately (the possible 
$N(\underline{n},D)$ contribution vanish). Taking into account the 
normalization (\ref{condT}) we get

\begin{eqnarray}
t(n_1,n_2,n_3<1,1,1,1,D)&=&\Df
{\overline{t}(n_1,n_2,n_3,D)}{\overline{t}(0,0,0,D)}\nonumber\\
&=&
\Df
{(2-D)_{(n_1+n_3)}
(2-D)_{(n_2+n_3)}
(\Df{D-1}{2})_{(-n_1-n_3)}
(\Df{D-1}{2})_{(-n_2-n_3)}}
{(-4)^{(n_1+n_2)}(-8)^{n_3}}
\nonumber\\
&&\sum_{k=0}^{[-n_3/2]}\Df{
(\Df{D-1}{2}-n_1-n_3)_{-k}
(\Df{D-1}{2}-n_2-n_3)_{-k}
(n_3)_{(-n_3-2k)}
}{
(\Df{3-D}{2})_{-k}
(\Df{1}{2})_{-k}
(-n_3-2k)!
}
\nonumber
\end{eqnarray}

For $n_3>1$ using integration by parts for $x_3$ in (\ref{tbar}) (which 
reduces to evaluation of $(n_3-1)^{th}$ derivative of $P^{D/2-2}$)
the $\overline{t}(n_1,n_2,n_3,D)$ can be reduced to 
a set of $\overline{t}(n_1,n_2,1,D)$ with different $n_1,n_2$.
Let us extract the $t(n_1,n_2,1,1,1,1,D)$ from $\overline{t}(n_1,n_2,1,D)$ 
according to the conditions (\ref{condT})

\begin{eqnarray}
t(n_1,n_2,1,1,1,1,D)=\Df{1}{\overline{t}(0,0,0,D)}
(\overline{t}(n_1,n_2,1,D)
-\overline{t}(0,0,1,D)N(n_1,n_2,1,1,1,1))
\label{t2}
\end{eqnarray}

One can calculate the $t(n_1,n_2,1,1,1,1,D)$
by direct use of the (\ref{t2})  expanding it for example in seria over
$D/2-2$, but we found more suitable to use the recurrence relations 
on $n_1,n_2$:

\begin{eqnarray}
t(n_1,n_2,1,1,1,1,D)&=&
-\Df{(D-2)^2}{4(D-3)(2n_1-D+2)}
(-\Df{1}{2}t(n_1-1,n_2-1,0,1,1,1,D-2)\nonumber\\
&&+t(n_1-2,n_2-1,1,1,1,1,D-2))\nonumber\\
&&-\Df{2(D-2)^2(11D-38)}{3(3D-10)(3D-8)(D-3)}
N(n_1,n_2,1,1,1,1,D)\label{rt1}\\
t(n_1,n_2,1,1,1,1,D)&=&
\Df{(n_1-n_2-1)}{(2n_1-D+2)}
t(n_1,n_2+1,0,1,1,1,D)\nonumber\\
&&+\Df{(2n_2-D+4)}{(2n_1-D+2)}
t(n_1-1,n_2+1,1,1,1,1,D)
\label{rt2}
\end{eqnarray}

With the help of (\ref{rt1}) the $n_1+n_2$ can be reduced to $-1,0,1$ 
and with the help of (\ref{rt2}) the $n_1-n_2$ can be reduced to $0,1$
(note that $t(n_1,n_2,1,1,1,1,D)=t(n_2,n_1,1,1,1,1,D)$).
Here at every recursion step the one integral reexpreses through
the other one plus rational over $D$, that is there is no 
"exponential reproduction". Then, the recursion acts separately
on variables $n_1+n_2$ and $n_1-n_2$. So, although the relations 
(\ref{rt1},\ref{rt2}) can be solved to explicit formulas,
this "safe" variant
of recursion is in this case the most effective way of calculations.

The relations (\ref{rt1},\ref{rt2}) are the simple example of the recurrence
relations with D-shifts, which can be derived in the following way.
Note that if 
$f^k(n_i,D)$ is a solution of (\ref{rr2}), then
$P({\bf I}^-+\mu_i)f^k(n_i,D-2)$ also is a solution.
Hence, if $f^k(n_i,D)$ is a complete set of solutions, then

\begin{eqnarray}
f^k(n_i,D)=\sum_n S^k_n(D)P({\bf I}^-+\mu_i)f^n(n_i,D-2)
\label{rrD}
\end{eqnarray}

\noindent
where the coefficients of mixing matrix $S$ depends only over $D$.
For the solutions (\ref{solution}) the matrix $S$ is unit matrix.
On the other hand, the desire to come to some specific set of master 
integrals leads to nontrivial mixing matrix and for the example considered 
above these coefficients are

\vspace{3mm}
\begin{tabular}{lll}
$S^n_n=-\Df{3}{64}\Df{(3D - 8)(3D - 10)}{(D - 4)^2}$
&$S^m_m=\Df{3}{16}\Df{(3D - 8)(3D - 10)}{(2D - 7)(2D - 9)}$&
$S^t_t=-\Df{(D - 2)^2}{4(D - 3)(D - 4)}$\\
$S^t_n=\Df{(11D - 38)(D - 2)^2}{32(D - 3)(D - 4)^2}$& 
\multicolumn{2}{l}{$S^n_t=S^t_m=S^m_t=S^m_n =S^n_m=0$}\\
\end{tabular}

\vspace{3mm}
To check the efficiency of this approach we 
evaluated, to 3 loops, the first 5 moments in the 
$z\equiv q^2/4m^2\to 0$ 
expansion of the QED photon vacuum polarization 
 \[\Pi(z)  =  \sum_{n>0} C_n\,z^n + {\rm O}(\alpha^4)\,,\]

The $C_n$ are expressed through approximately $10^5$ scalar 
integrals, but there is no necessary to evaluate these integrals separately.
Instead, we evaluated a few integrals of (\ref{solution}) type, but
with $P^{D/2-2}$ producted by a long polinomial in $x_i$.

After OS mass \cite{GBGS,BGS} and charge\cite{REC} renormalization,
we obtained the finite $D\rightarrow4$ limits
(the coefficients $C_1, C_2, C_3$ can be found in \cite{3l}):

\begin{eqnarray}
C_4 & = & \Bigl\{ N^2\left[ \Df{256}{693}                         \,\zeta_2
                          + \Df{2522821}{9437184}              \,\zeta_3
                        - \Df{129586264289}{143327232000})\right]\nonumber\\
     &  &{}     + N  \left[ \Df{160}{231} \left(1-\Df{8}{5}\ln2\right)\zeta_2
                          + \Df{1507351507033}{1651507200}        \,\zeta_3
                          - \Df{269240669884818833}{245806202880000}   \right]
                           \Bigr\}\frac{\alpha^3}{\pi^3} 
\nonumber\\
     &  &{}+\Df{51986}{127575}\,N\frac{\alpha^2}{\pi^2}
           +\Df{32}{693}       \,N\frac{\alpha  }{\pi  }\,,\nonumber\\
C_5 & = & \Bigl\{ N^2\left[ \Df{1024}{3003}                         
\,\zeta_2
                          + \Df{1239683}{3932160}                     \,\zeta_3
                          - \Df{512847330943}{556351488000})\right]\nonumber\\
     &  &{}     + N  \left[ \Df{640}{1001} \left(1-\Df{8}{5}\ln2\right)\zeta_2
                          + \Df{939939943788973}{190749081600}           \,\zeta_3
                          - \Df{360248170450504167133}{60837035212800000}    \right]
                           \Bigr\}\frac{\alpha^3}{\pi^3}  
\nonumber\\
     &  &{}+\Df{432385216}{1260653625}\,N\frac{\alpha^2}{\pi^2}
           +\Df{512}{15015}       \,N\frac{\alpha  }{\pi  }\,,\nonumber
\end{eqnarray}

\noindent
where we follow common practice \cite{BKT}, by allowing for
$N$ degenerate leptons. In pure QED, $N=1$; formally,
the powers of $N$ serve to count the number of electron loops.

The $N$ contribution of $C_4$ is in agreement with recent QCD 
calculations \cite{chet}, the $N^2$ part of $C_4$ and the 
$C_5$ are new.

The bare (non-renormalized) integrals were calculated for arbitrary $D$.
Calculations for $C_4$ were made on PC with 
Pentium-75 processor by REDUCE with 24Mbyte memory, within 
approximately 10 CPU hours. 
The most difficult diagrams for $C_5$ were calculated
on HP735 workstation.  

These results demonstrates a reasonable progress in comparison with common 
recursive 
approach. For example, the common way used in \cite{3l} demands  
several CPU hours on DEC-Alpha workstation to calculate full $D$ 
dependence of $C_2$ integrals, and further calculations became 
possible only after truncation in $(D/2-2)$. In the present approach the 
full $D$ calculations for $C_2$ demand about 5 minutes on PC.

\section{Conclusions}

The new approach suggested in this work allows to produce explicit 
formulas (\ref{solution}) for the solutions of the recurrence relations for 
3--loop vacuum integrals.
This formulas can be used for direct calculations and demonstrate a
high efficiency.
On the other hand, they produce a new type $D$-shifted recurrence 
relations (\ref{rrD}) for these integrals.
Finally, we hope that simple representation (\ref{rr2}) of the 
traditional recurrence relations which allows to obtain all these
results is not intrinsic for 3--loop vacuum case and 
generalization for multi--loop or/and non-vacuum case is possible.

\section{Acknowledgment}
I would like to thank D.Broadhurst for the possibility to use
his RECURSOR \cite{REC} which produced a lot of initial materials for 
investigating the structure of the solutions and V.Ilyin for drawing the 
attention to the problem and many fruitful discussions.

\end{document}